
\input harvmac

 %
\catcode`@=11
\def\rlx{\relax\leavevmode}                  
 %
 %
 %
\font\tenmib=cmmib10
\font\sevenmib=cmmib10 at 7pt 
\font\fivemib=cmmib10 at 5pt  
\font\tenbsy=cmbsy10
\font\sevenbsy=cmbsy10 at 7pt 
\font\fivebsy=cmbsy10 at 5pt  
\def\BMfont{\textfont0\tenbf \scriptfont0\sevenbf
                              \scriptscriptfont0\fivebf
            \textfont1\tenmib \scriptfont1\sevenmib
                               \scriptscriptfont1\fivemib
            \textfont2\tenbsy \scriptfont2\sevenbsy
                               \scriptscriptfont2\fivebsy}
\def\BM#1{\rlx\ifmmode\mathchoice
                      {\hbox{$\BMfont#1$}}
                      {\hbox{$\BMfont#1$}}
                      {\hbox{$\scriptstyle\BMfont#1$}}
                      {\hbox{$\scriptscriptstyle\BMfont#1$}}
                 \else{$\BMfont#1$}\fi}
 %
 %
 %
 %
\def\inbar{\vrule height1.5ex width.4pt depth0pt}
\def\sinbar{\vrule height1ex width.35pt depth0pt}
\def\ssinbar{\vrule height.7ex width.3pt depth0pt}
\font\cmss=cmss10
\font\cmsss=cmss10 at 7pt
\def\ZZ{\rlx\leavevmode
             \ifmmode\mathchoice
                    {\hbox{\cmss Z\kern-.4em Z}}
                    {\hbox{\cmss Z\kern-.4em Z}}
                    {\lower.9pt\hbox{\cmsss Z\kern-.36em Z}}
                    {\lower1.2pt\hbox{\cmsss Z\kern-.36em Z}}
               \else{\cmss Z\kern-.4em Z}\fi}
\def\Ik{\rlx{\rm I\kern-.18em k}}  
\def\IC{\rlx\leavevmode
             \ifmmode\mathchoice
                    {\hbox{\kern.33em\inbar\kern-.3em{\rm C}}}
                    {\hbox{\kern.33em\inbar\kern-.3em{\rm C}}}
                    {\hbox{\kern.28em\sinbar\kern-.25em{\sevenrm C}}}
                    {\hbox{\kern.25em\ssinbar\kern-.22em{\fiverm C}}}
             \else{\hbox{\kern.3em\inbar\kern-.3em{\rm C}}}\fi}
\def\IP{\rlx{\rm I\kern-.18em P}}
\def\IR{\rlx{\rm I\kern-.18em R}}
\def\Ione{\rlx{\rm 1\kern-2.7pt l}}
 %
 %

 %

\def\intem#1{\par\leavevmode%
              \llap{\hbox to\parindent{\hss{#1}\hfill~}}\ignorespaces}
 %


 %
\newskip\humongous \humongous=0pt plus 1000pt minus 1000pt   
\def\caja{\mathsurround=0pt}
\newif\ifdtup
 %
\def\eqalign#1{\,\vcenter{\openup2\jot \caja
     \ialign{\strut \hfil$\displaystyle{##}$&$
      \displaystyle{{}##}$\hfil\crcr#1\crcr}}\,}
 %

 %

 %

 %

 %

 %

 %
 %
 %
 %
\def\,{\hskip1.5pt}           
 %

\let\c=\chi

 %
 %
\def\Box{\sqcap\llap{$\sqcup$}}
\def\lapp{\lower.4ex\hbox{\rlap{$\sim$}} \raise.4ex\hbox{$<$}}
\def\gapp{\lower.4ex\hbox{\rlap{$\sim$}} \raise.4ex\hbox{$>$}}
\def\con{\ifmmode\raise.1ex\hbox{\bf*}
          \else\raise.1ex\hbox{\bf*}\fi}
\def\bo{{\raise.15ex\hbox{\large$\Box\kern-.39em$}}}

\def\dual{\relax\leavevmode\lower.9ex\hbox{\titlerms*}}

\let\8=\otimes
 %
 %
 %
 %

\let\2=\underline

 %
\def\dt#1{{\buildrel{\smash{\lower1pt\hbox{.}}}\over{#1}}}

\font\eightrm=cmr8
\def\6(#1){\relax\leavevmode\hbox{\eightrm(}#1\hbox{\eightrm)}}
\def\0#1{\relax\ifmmode\mathaccent"7017{#1}     
                \else\accent23#1\relax\fi}      
\def\7#1#2{{\mathop{\null#2}\limits^{#1}}}      
\def\5#1#2{{\mathop{\null#2}\limits_{#1}}}      
 %

 %

 %

 %

 %
\newbox\t@b@x
\def\rightarrowfill{$\m@th \mathord- \mkern-6mu
     \cleaders\hbox{$\mkern-2mu \mathord- \mkern-2mu$}\hfill
      \mkern-6mu \mathord\rightarrow$}
\def\tooo#1{\setbox\t@b@x=\hbox{$\scriptstyle#1$}%
             \mathrel{\mathop{\hbox to\wd\t@b@x{\rightarrowfill}}%
              \limits^{#1}}\,}
\def\leftarrowfill{$\m@th \mathord\leftarrow \mkern-6mu
     \cleaders\hbox{$\mkern-2mu \mathord- \mkern-2mu$}\hfill
      \mkern-6mu \mathord-$}
\def\froo#1{\setbox\t@b@x=\hbox{$\scriptstyle#1$}%
             \mathrel{\mathop{\hbox to\wd\t@b@x{\leftarrowfill}}%
              \limits^{#1}}\,}
 %
\def\frac#1#2{{#1\over#2}}
\def\frc#1#2{\relax\ifmmode{\textstyle{#1\over#2}} 
                    \else$#1\over#2$\fi}           
 %
\def\Claim#1#2#3{\bigskip\begingroup%
                  \xdef #1{\secsym\the\meqno}%
                   \writedef{#1\leftbracket#1}%
                    \global\advance\meqno by1\wrlabeL#1%
                     \noindent{\bf#2}\,#1{}\,:~\sl#3\vskip1mm\endgroup}

\def\QED{\rlx\hfill$\Box$\kern-7pt\raise3pt\hbox{$\surd$}\bigskip}
 %
 %

 %
\def\muthstrut{\vphantom1}
\def\mutrix#1{\null\,\vcenter{\normalbaselines\m@th
        \ialign{\hfil$##$\hfil&&~\hfil$##$\hfill\crcr
            \muthstrut\crcr\noalign{\kern-\baselineskip}
            #1\crcr\muthstrut\crcr\noalign{\kern-\baselineskip}}}\,}

 %
\def\YT#1#2{\vcenter{\hbox{\vbox{\baselineskip0pt\parskip=\medskipamount%
             \def\Box{$\sqcap\llap{$\sqcup$}$\kern-1.2pt}%
              \def\Z{\hfil\vskip-5.8pt}\lineskiplimit0pt\lineskip0pt%
               \setbox0=\hbox{#1}\hsize\wd0\parindent=0pt#2}\,}}}
\def\EU{\rlx\ifmmode \c_{{}_E} \else$\c_{{}_E}$\fi}
\def\TM{\rlx\ifmmode {\cal T_M} \else$\cal T_M$\fi}
\def\TW{\rlx\ifmmode {\cal T_W} \else$\cal T_W$\fi}
\def\CM{\rlx\ifmmode {\cal T\rlap{\bf*}\!\!_M}
             \else$\cal T\rlap{\bf*}\!\!_M$\fi}
\def\hm#1#2{\rlx\ifmmode H^{#1}({\cal M},{#2})
                 \else$H^{#1}({\cal M},{#2})$\fi}
\def\CP#1{\rlx\ifmmode\IP^{#1}\else\IP$^{#1}$\fi}
\def\cP#1{\rlx\ifmmode\IC{\rm P}^{#1}\else$\IC{\rm P}^{#1}$\fi}

\def\sll#1{\rlx\rlap{\,\raise1pt\hbox{/}}{#1}}
\def\Sll#1{\rlx\rlap{\,\kern.6pt\raise1pt\hbox{/}}{#1}\kern-.6pt}
%

 %
 %
\def\ie{\hbox{\it i.e.}}        

\def\CY{Calabi-\kern-.2em Yau}

\def\3{\ifmmode\ldots\else$\ldots$\fi}
\def\Z{\hfil\break\rlx\hbox{}\quad}
\def\3{\ifmmode\ldots\else$\ldots$\fi}
\def\?{d\kern-.3em\raise.64ex\hbox{-}}           
\def\9{\raise.43ex\hbox{-}\kern-.37em D}         

 %
 %

 %

 %

\def\NP#1{{\it Nucl.\,Phys.\,}{\bf#1\,}}
\def\PL#1{{\it Phys.\,Lett.\,}{\bf#1\,}}

\def\MPL#1{{\it Mod.\,Phys.\,Lett.\,}{\bf#1\,}}

\def\CMP#1{{\it Commun.\,Math.\,Phys.\,}{\bf#1\,}}

 %
 %
 %
\baselineskip=13.0861pt plus2pt minus1pt
\parskip=\medskipamount
\let\ft=\foot
\noblackbox
\def\SaveTimber{\abovedisplayskip=1.5ex plus.3ex minus.5ex
                \belowdisplayskip=1.5ex plus.3ex minus.5ex
                \abovedisplayshortskip=.2ex plus.2ex minus.4ex
                \belowdisplayshortskip=1.5ex plus.2ex minus.4ex
                \baselineskip=12pt plus1pt minus.5pt
 \parskip=\smallskipamount
 \def\ft##1{\unskip\,\begingroup\footskip9pt plus1pt minus1pt\setbox%
             \strutbox=\hbox{\vrule height6pt depth4.5pt width0pt}%
              \global\advance\ftno by1\footnote{$^{\the\ftno)}$}{##1}%
               \endgroup}
 \def\listrefs{\footatend\vfill\immediate\closeout\rfile%
                \writestoppt\baselineskip=10pt%
                 \centerline{{\bf References}}%
                  \bigskip{\frenchspacing\parindent=20pt\escapechar=` %
                   \rightskip=0pt plus4em\spaceskip=.3333em%
                    \input refs.tmp\vfill\eject}\nonfrenchspacing}}
 %
\def\Afour{\ifx\answ\bigans
            \hsize=16.5truecm\vsize=24.7truecm
             \else
              \hsize=24.7truecm\vsize=16.5truecm
               \fi}
\catcode`@=12

\let\al=\alpha

\def\eqaligntwo#1{\,\vcenter{\openup2\jot \caja
     \ialign{\strut \hfil$\displaystyle{##}$&
                         $\displaystyle{{}##}$\hfil&
                          $\displaystyle{{}##}$\hfil\crcr#1\crcr}}\,}

 \def\Afour{\hsize=16.5truecm\vsize=24.7truecm}

\baselineskip=12pt

\Title{\vbox{\baselineskip12pt \hbox{IASSNS-HEP-93/92}}}
       {\vbox{\centerline{Landau-Ginzburg Orbifolds, Mirror Symmetry}
             \vskip10pt
             \centerline{ and the Elliptic Genus}}}

\centerline{Per Berglund\footnote{$^{\dag}$}
{Email: berglund@guinness.ias.edu}
 and M{\aa}ns Henningson\footnote{$^{\ddag}$}
{Email: mans@guinness.ias.edu}} \vskip4mm
 \centerline{\it School of Natural Science} \vskip 0mm
 \centerline{\it Institute for Advanced Study}      \vskip 0mm
 \centerline{\it Olden Lane}               \vskip 0mm
 \centerline{\it Princeton, NJ 08540}                \vskip 0mm
 \vfill

\centerline{ABSTRACT}\vskip2mm
\vbox{\narrower\narrower\baselineskip=12pt\noindent
We compute the elliptic genus for arbitrary
two dimensional $N=2$ Landau-Ginzburg orbifolds. This is
used to search for possible mirror pairs of such models. We show that if two
Landau-Ginzburg models are conjugate to each other in a certain sense, then to
every orbifold of the first theory corresponds an orbifold of the second theory
with the same elliptic genus (up to a sign) and with the roles of the chiral
and anti-chiral rings interchanged. These orbifolds thus constitute a possible
mirror pair. Furthermore, new pairs of conjugate models may be obtained by
taking the product of old ones. We also give a sufficient (and possibly
necessary) condition for two models to be conjugate, and show that it is
satisfied by the mirror pairs proposed by one of the authors and~H\"ubsch.
}

\Date{\vbox{
       \line{12/93 \hfill}}}

\noblackbox
\lref\rDixon{For a review and references, see L.~Dixon, in {\it
     Superstrings, Unified Theories and Cosmology 1987},
     eds.~G.~Furlan et al.\ (World Scientific, Singapore, 1988).}

\lref\rCHSW{P.~Candelas, G.~Horowitz, A.~Strominger and E.~Witten,
      \NP{B258} (1985) 46.}

\lref\rMPR{P.~Candelas, M.~Lynker and R.~Schimmrigk,
      \NP{B341} (1990) 383.}

\lref\rGP{B.~R.~Greene and M.~R.~Plesser, \NP{B338} (1990) 15.}

\lref\rVW{C.~Vafa and N.P.~Warner, \PL{218B} (1989) 377\semi
          E.Martinec, \PL{B217} (1989) 431.}

\lref\rVS{C.~Vafa, \MPL{A4} (1989) 1169.}

\lref\rAR{A.~Klemm and R.~Schimmrigk~, ``Landau-Ginzburg String Vacua'',
       {\it CERN preprint} CERN-TH 6459/92\semi
          M.~Kreuzer and H.~Skarke, \NP{B388} (1992) 113.}

\lref\rBH{P.~Berglund and T.~H\"ubsch, in {\it Essays on Mirror Manifolds},
S.T.~Yau ed. (International Press, Hong Kong 1992); \NP{B393} (1993) 377.}

\lref\rIV{K.~Intrilligator and C.~Vafa, \NP{B339} (1990) 95.}

\lref\rASY{O.~Aharony, A.N.~Schellekens and S.~Yankielowicz,
``Charge Sum Rules in $N=2$ Theories'', NIKHEF-H/93-27 and
Tel-Aviv University report TAUP-2123-93.}

\lref\rKYY{T.~Kawai, Y.~Yamada and S.-K.~Yang,
``Elliptic Genera and $N=2$ Superconformal Field Theory'',
KEK-TH-362.}

\lref\rFY{P.~Di~Francesco and S.~Yankielowicz,
\NP{B409} (1993) 186.}

\lref\rMH{M.~Henningson, ``N=2 Gauged WZW models and the Elliptic Genus'',
IAS preprint IASSNS-HEP-39/93, \NP{B} to appear.}

\lref\rAFY{O.~Aharony, P.~Di~Francesco and S.~Yankielowicz,
``Elliptic Genera and the Landau-Ginzburg Approach to $N=2$ Orbifolds'',
Saclay preprint SPhT 93/068 and Tel-Aviv University report TAUP 2069-93.}

\lref\rWitten{E.~Witten, ``On the Landau-Ginzburg Description of $N=2$
minimal models'', IAS preprint IASSNS-HEP-93/10.}

\lref\rLVW{W.~Lerche, C.~Vafa and N.P.~Warner, \NP{B324} (1989) 427.}

\lref\rVT{C.~Vafa, \NP{B273} (1986) 592.}

\lref\rGQ{D.~Gepner and Z.~Qiu, \NP{B285} (1987) 423.}

\lref\rSW{A. Schellekens and N. Warner, \PL{177B} (1986) 317\semi
A. Schellekens and N. Warner, \PL{181B} (1986) 339\semi
K. Pilch, A. Schellekens and N. Warner, \NP{B287} (1987) 317.}

\lref\rWittenCMP{E. Witten, Comm. Math. Phys. {\bf 109} (1987) 525 \semi
E. Witten, in {\it Elliptic Curves and Modular Forms in Algebraic Topology} ed.
P. Landwebber, (Springer Verlag 1988).}

\lref\rGSW{M.B. Green, J.H. Schwarz and E. Witten, {\it Superstring
Theory} (Cambridge monographs on mathematical physics 1987).}

\lref\rWW{E.T.~Whittaker and G.N.~Watson, {\it A Course of Modern Analysis},
(Cambridge University Press 1958).}

\lref\rBatdual{V.~Batyrev, ``Dual polyhedra and mirror symmetry for Calabi-Yau
hypersurfaces in toric varieties'', Preprint, November 18, 1992.}

\lref\rCOK{P.~Candelas, X.~de~la~Ossa and S.~Katz, ``Mirror Symmetry for
Calabi-Yau Hypersurfaces in Weighted $\IP^4$ and an Extension of
Landau-Ginzburg Theory'', in preparation.}

\lref\rMax{M.~Kreuzer and H.~Skarke, ``All Abelian symmetries of
Landau-Ginzburg potentials'', CERN-TH-6705-92.}

\lref\rArnold{V.I.~Arnold, S.M.~Gusein-Zade and A.N.~Varchenko, {\it
     Singularities of Differentiable Maps, Vol.~I}~ (Birkh\"auser,
     Boston, 1985).}

\lref\rCdGP{P.~Candelas, X.~ de la Ossa, P.~Green and L.~Parkes,
       \NP{B359}~(1991)~21.}

\lref\rAGM{P.S.~Aspinwall, B.R.~Greene and D.R.~Morrison, ``\CY\ moduli
      space, mirror manifolds and spacetime topology change'', IAS preprint
      IASSNS-HEP-93/38.}

\lref\rDHVW{L.~Dixon, J.~Harvey, C.~Vafa and E.~Witten, \NP{B261} (1985)
678\semi
L.~Dixon, J.~Harvey, C.~Vafa and E.~Witten, \NP{B274} (1986) 285.}

\lref\rKS{Y.~Kazama and H.~Suzuki, \NP{B321} (1989) 232\semi
Y.Kazama and H. Suzuki, \PL{216B}~(1989)~112.}

\lref\rks{M.~Kreuzer and H.~Skarke, \CMP{150} (1992) 137.}

\newsec{Introduction}\noindent
Conformal field theories with $N=2$ supersymmetry have been much
studied in the last few years. The original motivation was to construct
vacua of the heterotic string which exhibit $N=1$ space-time supersymmetry.
This requires two right-moving supersymmetries on the world-sheet, but most
investigations have focused on the less general case of models with two
left-moving supersymmetries as well, so called $(2,2)$ models~\rDixon.
 String theory
leads us to consider models with conformal anomaly $\hat c=3$, but $N=2$
superconformal field theories are interesting in their own right without
reference to any particular value of $\hat c$.

Known examples of $N=2$ superconformal field theories include sigma models on
Calabi-Yau targetspaces of complex dimension $\hat c$ (which must be an integer
in this case)~\rCHSW\ and Kazama-Suzuki coset models~\rKS. The latter include
the so called $N=2$ minimal models as a special case. Another interesting class
of models are the so called $N=2$ Landau-Ginzburg models, \ie\ supersymmetric
quantum field theories, the superpotentials of which have a degenerate critical
point. Although not conformally invariant as they are usually written, they are
believed to flow to a fixed point in the infrared~\rVW. Furthermore, the
superpotential is believed to be an invariant under this renormalization group
flow. Properties of these models that only depend on the superpotential may
thus be studied away from the fixed point, where calculations are easier to
perform.

{}From an $N=2$ model, conformally invariant or not, with some symmetry group
$H$, we may construct new models by a process usually referred to as
orbifolding~\rDHVW. This means that we identify field configurations
modulo the action of the group $H$. Another way of expressing this is to
allow for twisted boundary conditions along the cycles of the worldsheet.

In some cases, the superconformal field theory to which a particular
Landau-Ginzburg theory flows has been conjectured. Until recently, the main
support for these conjectures came from comparing the chiral and anti-chiral
rings in the respective theories~\rLVW, but this only probes the zero-modes of
the theories and gives little information about `stringy' effects.
Such information
would be provided by the partition function $Z(q,\gamma_L,\gamma_R) = {\rm Tr}
(-1)^F q^{L_0} {\bar q}^{\bar L_0} \exp (i\gamma_L J_0 + i\gamma_R {\bar
J}_0)$.
Here, $L_0$ ($\bar{L}_0$) and $J_0$ (${\bar J}_0$) are the energy and $U(1)$
charge operator respectively of the left-moving (right-moving) $N=2$ algebra.
However,
although the partition function of for example the minimal models and their
orbifolds is known, it is not effectively calculable for a general $N=2$
theory.
The situation is much better for the elliptic genus \rSW\ \rWittenCMP,
which is simply the
restriction of the partition function to $\gamma_R=0$, \ie\ $Z(q,\gamma,0)$.
The reason that it can be computed in many practical cases is related to the
fact that it has an interpretation as an index of the $N=1$ right-moving
supercharge. It is therefore invariant under smooth deformations of the theory
which preserve an $N=1$ right-moving supersymmetry. This invariance was
recently used by Witten~\rWitten\ to calculate the elliptic genera of certain
Landau-Ginzburg models which are believed to flow to the minimal models. The
results were compared to the elliptic genera of the minimal models, calculated
from the known characters of the $N=2$ discrete series representations,
in~\refs{\rWitten,\rFY}.
The elliptic genus has also been calculated for the minimal models in
their formulation as $SU(2)/U(1)$ Kazama-Suzuki coset models in~\rMH.
For other applications of the elliptic genus, see~\refs{\rAFY,\rKYY,\rASY}.

Mirror symmetry is an intriguing property of the space of $N=2$ theories. At
the level of conformal field theory, it could be formulated as an isomorphism
between two theories, amounting to a change of sign of, say, the left-moving
$U(1)$ generator. Since the sign of the charge is a pure convention, this may
seem rather trivial, but mirror symmetry leads to some remarkable conclusions.
For example, in the case of Calabi-Yau sigma-models, mirror symmetry states
that target spaces of different topological type give rise to equivalent
space-time physics. Mirror symmetry was first conjectured because it seemed
unnatural that the conventional choice of sign of the $U(1)$ generators should
have any independent meaning~\refs{\rDixon,\rLVW}. Following a previous
observation by Gepner and Qiu~\rGQ\ that the $\ZZ_k$ orbifold of the $A_k$
minimal model is equivalent to the $A_k$ model itself, Greene and Plesser~\rGP\
gave the first explicit example of mirror symmetry by considering orbifolds of
tensor products of minimal models. Although there are many indications that
mirror symmetry is indeed a
symmetry of the space of $(2,2)$
theories~\refs{\rMPR,\rks,\rBH,\rAR,\rMax,\rBatdual,\rCOK,\rCdGP,\rAGM},
the Greene-Plesser
construction is the only one which has been rigorously proved at the level of
conformal field theory.

Using the conjectured Landau-Ginzburg formulation of the minimal models, the
Greene-Plesser construction may be thought of as an equivalence between
Landau-Ginzburg orbifolds. It is then natural to ask whether there are other
such mirror pairs of Landau-Ginzburg orbifolds. In~\rBH\, two classes of
examples were proposed which generalize the minimal models, and some
comparisons of the chiral and anti-chiral rings were made which supported the
conjecture of mirror symmetry.

Most of the interest in mirror symmetry has been concentrated on models with
integer $\hat c$, in particular $\hat c = 3$, which allow a geometric
interpretation in terms of a sigma-model with Calabi-Yau target-space. Also,
one usually imposes a projection on states of integral $U(1)$ charge, as is
necessary for the consistency of the string theory interpretation. We would
like to stress the two-dimensional point of view, however.
Indeed, the examples
in~\rBH\ seem to work equally well for models with any $\hat c$ and fractional
$U(1)$ charges.
In fact our computation may indicate that mirror symmetry is a property of
not only $N=2$ superconformal field theories but two dimensional $N=2$
quantum field theories in general.
Of course, we hope that a study of mirror symmetry in this
broader context will teach us about phenomena that are relevant to string vacua
as well.

In this article, we will use the elliptic genus as a tool for studying
mirror symmetry in the context of Landau-Ginzburg orbifolds. Namely, since the
spectrum of an $N=2$ theory is symmetric under charge conjugation, the elliptic
genus is invariant under $J_0 \rightarrow -J_0$, \ie\ $Z(q, \gamma, 0) =  Z(q,
-\gamma, 0)$. The elliptic genera of two models that constitute a mirror pair
must therefore be equal (up to a sign, which could be thought of as arising
from different normalizations of the path integral measures of the two
theories). Of course, the mere equality of the elliptic genera is by no means a
proof of mirror symmetry. In particular, because of the index interpretation of
the elliptic genus, it is independent of the exact point in the moduli space of
$N=2$ theories that we are considering. Equality of the elliptic genera of two
models is therefore only an indication that some point in the component of
moduli space in which the first model lies might be the mirror partner of some
point
in
the component of moduli space of the second model. However, in this article,
we will for brevity
refer to two models with the same elliptic genus as constituting a mirror pair.

This article is organized as follows: In section~2, we calculate the elliptic
genus for an arbitrary Landau-Ginzburg orbifold. In section~3 we calculate the
Poincar\'{e} polynomial of these theories by taking the $q \rightarrow 0$ limit
of the elliptic genus. We also review the proof, first given by Francesco and
Yankielowicz~\rFY, that the Poincar\'{e} polynomials completely determines the
elliptic genus in the case of Landau-Ginzburg orbifolds. In section~4, we
discuss a plausible scenario for mirror symmetry between Landau-Ginzburg
orbifolds. Our main results are that if two Landau-Ginzburg models are
conjugate to each other in a certain sense, then every orbifold of the first
model is the mirror partner of some orbifold of the second model. Furthermore,
given two models and their conjugate partners, the product models are conjugate
to each other. Using the previous result that the Poincar\'{e} polynomial
determines the elliptic genus, we also give a sufficient condition for two
models to be each others conjugate.
Although we have no proof of the necessity of this condition, it covers
all cases of mirror symmetry between Landau-Ginzburg orbifolds that we know of.
Indeed, in section~5, we show that the models proposed in~\rBH\ satisfy this
condition. We conjecture that all pairs of conjugate Landau-Ginzburg models may
be obtained by taking products of these models.

\newsec{Landau-Ginzburg orbifolds and the elliptic genus}
\noindent
In this section, we will review some facts about $N=2$ Landau-Ginzburg models
and their orbifolds. In particular, we will show how their elliptic genera can
be calculated~\refs{\rWitten,\rFY,\rAFY-\rASY}.

The action of a $(2,2)$ Landau-Ginzburg model can be written in superspace as
\eqn\eaction{
S = \int d^2 z \, d^4 \theta \, K(X_i, \bar{X}_i) + \epsilon \int d^2 z \, d^2
\theta \, W(X_i) + {\rm c.c.},
}
where $X_i$ for $i \in N$ are complex chiral superfields with component
expansions
\eqn\esuperf{
X_i = x_i + \theta_+ \psi_i^+ + \theta_- \psi_i^- + \theta_+ \theta_- F_i.
}
Here $N$ is a set that indexes the fields $X_i$
and the number of fields is denoted $|N|$.
The superpotential $W$ is a holomorphic and
quasi-homogeneous function of the $X_i$, \ie\ it should be possible to assign
some weights $k_i \in \ZZ$ to the fields $X_i$ for $i \in N$ and a degree
of homogeneity $D \in \ZZ$ to $W$ such that
\eqn\escale{
W(\lambda^{k_i} X_i) = \lambda^D W(X_i).
}
It will prove convenient to also introduce the charges
\eqn\eXXX{
q_i = \frac{k_i}{D} \;\; {\rm for} \;\; i \in N.
}
The model defined by~\eaction\ is believed to flow to a conformally
invariant model in the infrared. Under this renormalization group flow, the
K\"ahler potential $K(X_i, \bar{X}_i)$ will get renormalized in some
complicated way, but there are strong reasons to believe that the
superpotential $W(X_i)$ is an invariant of the flow~\rVW.

In general, $W$ will be invariant under a discrete, abelian group $G$ of phase
symmetries. The fields $X_i$ transform in some representations $R_i$ under $G$.
In the following, we denote the set of representations $\{ R_i \}$ for $i \in
N$ collectively as $R$. A representation $R_i$ of $G$ is specified by a
function $R_i(g) = \exp (i 2 \pi \theta_i(g))$ defined for $g \in G$, which
fulfils $R_i(g_1 g_2) = R_i(g_1) R_i(g_2)$ for all $g_1, g_2 \in G$. The
invariance of $W$ means that
\eqn\eXXX{
W(R_i(g) X_i) = W(X_i) \;\; {\rm for} \;\; g \in G.
}
By taking $\lambda = \exp i 2 \pi/D$ in~\escale, we see that $G$ will
always contain an element $q$ such that $R_i(q) = \exp i 2 \pi q_i$ for $i \in
N$. From a theory invariant under some symmetry group $H$, we may construct a
new theory by taking the $H$ orbifold of the original theory, \ie\ by modding
out by the action of $H$. In our case, $H$ could be any subgroup of the group
$G$ of phase symmetries of $W$. We will denote the theory thus obtained as
$W/H$~\ft{In general $q\notin H$ and hence $W/H$ is not a valid superstring
vacuum, even if $\hat c = 3$.}.

Our object is to calculate the elliptic genus of $W/H$. This calculation is
feasible because of the invariance of the elliptic genus under deformations of
the theory which preserve the right-moving supersymmetry~\rWitten.
We may therefore
smoothly turn off the superpotential interactions by letting $\epsilon
\rightarrow 0$, which turns the model into a free field theory. (We take $K$ to
be the K\"ahler potential of $\IC^{|N|}$ with the flat metric.)
The elliptic genus of the model is
determined by the set of representations $R$ and the group $H$ and may be
denoted as $Z[R/H]$, where we have suppressed the dependence on $\gamma$ and
$q$. Being essentially a genus one correlation function, it may be written as
\eqn\epartition{
Z[R/H] = \frac{1}{|H|} \sum_{h_a, h_b \in H} Z[R](h_a, h_b).
}
Here $Z[R](h_a, h_b)$ denotes the contribution from field configurations
twisted by $h_a$ and $h_b$ around the $a$ and $b$ cycles of the torus
respectively, and $|H|$ is the number of elements of $H$.
In the free field
limit, the contribution from each twist sector is a product of contributions
from each of the fields $X_i$ in the theory, \ie\
\eqn\eZsector{
Z[R](h_a, h_b) = \prod_{i \in N} Z[R_i](h_a, h_b).
}
In order  to calculate $Z[R_i](h_a, h_b)$, it is essential to know the charges
of the components of the superfield $X_i$ under the $U(1)$ symmetry which is
part of the left-moving $N=2$ superconformal algebra. A by now standard
calculation gives the charges of the component fields $x_i$, $\psi^+_i$ and
$\psi^-_i$ in~\esuperf\ as $q_i$, $q_i$ and $q_i-1$ respectively after the
path integral over the auxiliary fields $F_i$ has been carried out~\rWitten.
In a sector
twisted by $h_a$ along the `space' direction, the mode numbers of $X_i$ and
$\bar{X}_i$ are given modulo $\ZZ$ by $\theta_i(h_a)$ and $-\theta_i(h_a)$
respectively. With a twist of $h_b$ along the `time' direction, we should make
an insertion of $(-1)^F R_i(h_b) \exp (i 2 \pi \gamma J_0)$. The contribution
from the fermionic components of $X_i$ in this twist-sector is
\eqn\eXXX{
\eqalign{\prod_n^\infty &(1- \bar{q}^{n + \theta_i(h_a)}
R_i(h_b) e^{i 2 \pi q_i
\gamma}) (1- \bar{q}^{n - \theta_i(h_a)} R_i^{-1}(h_b) e^{- i 2 \pi q_i
\gamma}) \cr
\times &(1- q^{n + \theta_i(h_a)} R_i(h_b) e^{i 2 \pi (q_i-1)
\gamma}) (1- q^{n - \theta_i(h_a)} R_i^{-1}(h_b) e^{- i 2 \pi (q_i-1) \gamma}),
}}
where the four factors come from the modes of $\psi_i^+$, $\bar{\psi}_i^+$,
$\psi_i^-$ and $\bar{\psi}_i^-$ respectively. The contribution from, for
example, the leftmoving part of $x_i$ is
\eqn\eXXX{
\prod_n^\infty \sum_{s = 0}^\infty (\bar{q}^{n + \theta_i(h_a)} R_i(h_b) e^{i 2
\pi q_i \gamma})^s = \prod_n^\infty (1 - \bar{q}^{n + \theta_i(h_a)} R_i(h_b)
e^{i 2 \pi q_i \gamma})^{-1}.
}
Similarly, the total contribution from the left- and right-moving parts of
$x_i$ and $\bar{x}_i$ is
\eqn\eXXX{\hbox{}\mkern40mu
\eqalign{\prod_n^\infty &\left( (1 - \bar{q}^{n + \theta_i(h_a)}
R_i(h_b) e^{i 2 \pi
q_i \gamma}) (1- \bar{q}^{n - \theta_i(h_a)} R_i^{-1}(h_b) e^{- i 2 \pi q_i
\gamma})\right)^{-1} \cr
\times &\left( (1- q^{n + \theta_i(h_a)} R_i(h_b) e^{i 2 \pi q_i
\gamma}) (1- q^{n - \theta_i(h_a)} R_i^{-1}(h_b) e^{- i 2 \pi q_i \gamma})
\right)^{-1} .
}}
Putting together the fermionic and bosonic parts and multiplying both the
numerator and the denominator by $\prod_n^\infty (1 - q^n)$, we get
\eqn\eXXX{
\prod_n^\infty \frac{(1 {-} q^n) (1{-} q^{n {+} \theta_i(h_a)}
R_i(h_b) e^{i 2 \pi
(q_i{-}1) \gamma}) (1{-} q^{n {-} \theta_i(h_a)} R_i^{-1}(h_b)
e^{{-} i 2 \pi (q_i{-}1)
\gamma})}{(1 {-} q^n) (1{-} q^{n {+} \theta_i(h_a)} R_i(h_b)
e^{i 2 \pi q_i \gamma})
(1{-} q^{n {-} \theta_i(h_a)} R_i^{-1}(h_b) e^{- i 2 \pi q_i \gamma})}.
}
In this expression, a prefactor that depends on the energy and transformation
properties under $G$ of the vacuum in this twist-sector is missing. Also, we
have been somewhat cavalier in our treatment of the lower limits of the
products over $n$. We will come back to these points shortly. For the moment,
we note that the numerator and the denominator in our expression resemble
$\Theta_1((1-q_i) \gamma - \theta_i (h_b) - \tau \theta_i (h_a) | \tau)$ and
$\Theta_1(q_i \gamma + \theta_i (h_b) + \tau \theta_i (h_a) | \tau)$
respectively, where the Jacobi $\Theta_1$ function is defined by~\rGSW\
\eqn\eXXX{
\Theta_1(\nu | \tau) = i q^{1/8} e^{-i \pi \nu} (1 - e^{i 2 \pi \nu}) \prod_{n
= 1}^\infty (1 - q^n) (1 - q^n e^{i 2 \pi \nu}) (1 - q^n e^{- i 2 \pi \nu})
}
and $q = \exp (i 2 \pi \tau)$. However, the functions $\theta_i(h)$ are only
defined modulo $\ZZ$. We must make sure that the contribution to the
elliptic genus is well-defined, \ie\ it should only depend on
$R_i(h_a) = \exp
(i 2 \pi \theta_i(h_a))$ and $R_i(h_b)$. From the double quasi-periodicity
properties of the Jacobi $\Theta_1$ function~\rGSW,
\eqn\eXXX{
\eqalign{
\Theta_1(\nu + 1 | \tau) & =  - \Theta_1(\nu | \tau)  \cr
\Theta_1(\nu + \tau | \tau) & =  - e^{- i \pi (\tau + 2 \nu)} \Theta_1(\nu |
\tau) ,
}}
it follows that
\eqn\ethreesix{
Z[R_i](h_a, h_b) = e^{-i 2 \pi \gamma \theta_i(h_a)} \frac{\Theta_1((1-q_i)
\gamma - \theta_i(h_b) - \tau \theta_i(h_a) | \tau)}{\Theta_1(q_i \gamma +
\theta_i (h_b) + \tau \theta_i (h_a) | \tau)}
}
has the correct properties, \ie\ it is invariant under $\theta_i(h_a)
\rightarrow \theta_i(h_a)+1$ and $\theta_i(h_b) \rightarrow \theta_i(h_b)+1$.
Furthermore, if we define $L$ as the smallest integer such that $g^L$ is the
identity element of $G$ for $g \in G$, then
\eqn\eXXX{\hbox{}\mkern-20mu
\eqaligntwo{
Z[R_i](h_a, h_b) &\rightarrow (-1)^{(1 {-} 2 q_i) L} Z[R_i](h_a, h_b)
\;  &{\rm as} \;
(\gamma, \tau) \rightarrow (\gamma {+} L, \tau) \cr
Z[R_i](h_a, h_b) &\rightarrow (-1)^{(1 {-} 2 q_i) L}
e^{- i \pi (1 {-} 2 q_i) (L^2
\tau {+} 2 L \gamma)} Z[R_i](h_a, h_b)
\; &{\rm as} \; (\gamma, \tau) \rightarrow
(\gamma {+} L \tau, \tau).
}}
Next, we use the modular transformation properties of the Jacobi $\Theta_1$
function~\rGSW,
\eqn\eXXX{
\eqalign{
\Theta_1(\nu | \tau + 1) &= e^{i \pi /4} \Theta_1(\nu | \tau) \cr
\Theta_1(\nu / \tau | - 1/\tau) & = (-i \tau)^{1/2} e^{i \pi \nu^2 / \tau}
\Theta_1(\nu | \tau),
}}
to show that
\eqn\eXXX{\hbox{}\mkern20mu
\eqaligntwo{
Z[R_i](h_a, h_b) &\rightarrow Z[R_i](h_a, h_a h_b)
\qquad & {\rm as} \qquad (\gamma,
\tau) \rightarrow (\gamma, \tau + 1) \cr
Z[R_i](h_a, h_b) &\rightarrow e^{i \pi (1 - 2 q_i) \gamma^2/\tau} Z[R_i](h_b,
h_a^{-1})
\qquad & {\rm as} \qquad (\gamma, \tau)
\rightarrow (\gamma / \tau, -1 / \tau).
}}
The complete elliptic genus, given by~\epartition\ with~\eZsector, thus
transforms as
\eqn\emoduli{\hbox{}\mkern25mu
\eqaligntwo{
Z[R/H] &\rightarrow (-1)^{\hat{c} L} Z[R/H] \qquad & {\rm as} \qquad
(\gamma, \tau) \rightarrow (\gamma + L, \tau) \cr
Z[R/H] &\rightarrow (-1)^{\hat{c} L} e^{- i \pi \hat{c} (L^2 \tau + 2 L
\gamma)} Z[R/H] \qquad & {\rm as} \qquad
 (\gamma, \tau) \rightarrow (\gamma + L \tau,\tau) \cr
Z[R/H] &\rightarrow Z[R/H] \qquad & {\rm as} \qquad
(\gamma, \tau) \rightarrow (\gamma,\tau + 1) \cr
Z[R/H] &\rightarrow e^{i \pi \hat{c} \gamma^2 / \tau} Z[R/H]
\qquad & {\rm as} \qquad(\gamma, \tau)
\rightarrow (\gamma / \tau, - 1 / \tau) ,
}}
where $\hat{c} = \sum_{i \in N} (1 - 2 q_i)$ is the central charge of $N=2$
superconformal algebra~\rVW. These double quasi-periodicity and modular
transformation properties justify our Ansatz~\ethreesix. However, there are
other possibilities of constructing an elliptic genus which transforms in the
same way. Depending on the group $H$, there might be subsets of $H \times H$
which are closed under the action $(h_a, h_b) \rightarrow (h_a, h_a h_b)$ and
$(h_a, h_b) \rightarrow (h_b, h_a^{-1})$ of the modular group. For every such
subset there is a term, obtained by summing $Z[R](h_a, h_b)$ over the subset,
which may be added to the elliptic genus with an arbitrary coefficient without
destroying the above transformation properties. Modular covariance on
world-sheets of genus two imposes some restrictions on these
coefficients. The
remaining freedom corresponds to a choice of discrete torsion~\rVT. (Modular
covariance on higher genus world-sheets gives no further restrictions, since
such world-sheets can be obtained by
 sewing together world-sheets of lower genus~\rVT.)
In this paper, however, we will
always assume that the elliptic genus is given by the canonical
choice~\epartition\
with~\eZsector\ and~\ethreesix. (The simplest example of a
subset of $H \times H$ which is closed under modular transformations only
consists of $(e, e)$, where $e$ is the identity element of $H$. The
corresponding term $Z[R](e,e)$ obviously transforms correctly, being the
elliptic genus of the original Landau-Ginzburg theory before orbifolding.)

\newsec{The Poincar\'{e} polynomial and equality of elliptic genera}\noindent
In this section, we will define the Poincar\'{e} polynomial as a particular
limit of the elliptic genus. It turns out that the Poincar\'{e} polynomial
completely determines the elliptic genus in the case of Landau-Ginzburg
orbifolds. This property is of great value in comparing the elliptic genera of
different models.

The Poincar\'{e} polynomial of an $N=2$ superconformal field theory is the
$\tau \rightarrow i \infty$ (or equivalently $q \rightarrow 0$) limit of the
elliptic genus. It is a polynomial in $t^{1/D}$, where $t = \exp (i 2 \pi
\gamma)$ and $D$ is some integer (defined by~\escale\ for the case of
Landau-Ginzburg orbifolds). It can be seen as the generating function for the
charges of the Ramond sector ground states under the $U(1)$ symmetry of the
left-moving $N=2$ algebra. From our results in the last section, we see that
the Poincar\'{e} polynomial for a Landau-Ginzburg orbifold may be written as a
sum over contributions from different twist-sectors;
\eqn\efourthree{
P[R/H] = \frac{1}{|H|} \sum_{h_a, h_b \in H} P[R](h_a, h_b).
}
Furthermore, the contribution from each twist-sector is a product of the
contributions from each of the fields;
\eqn\efourfour{
P[R](h_a, h_b) = \prod_{i \in N} P[R_i](h_a, h_b).
}
Finally, by taking the $\tau \rightarrow i \infty$ limit of~\ethreesix, we
find that
\eqn\efourfive{
P[R_i](h_a, h_b) = \left\{
\eqaligntwo{
&- t^{1/2 {-} [\theta_i(h_a)]} \qquad & {\rm for} \; R_i(h_a) \neq 1 \cr
&t^{-1/2} (t^{q_i} R_i(h_b) {-} t)(1 {-} t^{q_i} R_i(h_b))^{-1}
\qquad & {\rm for} \; R_i(h_a) = 1 } \right. .
}
Here, $[x]$ denotes the fractional part of $x$, \ie\ $[x] = x {\;\; \rm mod
\;\;} \ZZ$ and $0 \leq [x] < 1$. Note that if $R_i(h_a) = 1$, or equivalently
$[\theta_i(h_a)] = 0$, then the field $X_i$ is left untwisted by the
transformation $h_a$. It will prove convenient to introduce
\eqn\eptw{
P^{\rm tw}[R](g) = \prod_{i \in N} (- t^{1/2})^{{\rm tw}[R](g)} \, t^{-\sum_{i
\in N} [\theta_i(g)]}
}
and
\eqn\epinv{
P^{\rm inv}[R_i](g) = t^{-1/2} (t^{q_i} R_i(g) - t)(1 - t^{q_i}
R_i(g))^{-1},
}
where ${\rm tw}[R](g)$ denotes the number of fields that are twisted by $g$. We
may then write
\eqn\efourtw{
P[R](h_a, h_b) = P^{\rm tw}[R](h_a) \, \prod_{R_i(h_a) = 1} P^{\rm
inv}[R_i](h_b).
}

Incidentally, there is a generalized Poincar\'{e} polynomial, which is
sensitive to the charges of the Ramond sector ground states under both the
left- and right-moving $U(1)$ symmetry. Since mirror symmetry acts by reversing
the sign of one of the $U(1)$ charges
and leaving the other unaffected, this generalized Poincar\'{e} polynomial is
useful to check whether two models might be each others mirror partners rather
than being completely equivalent theories. It may be defined as~\rLVW\
\eqn\eXXX{
P(t, \bar{t}) = {\rm Tr} (t^{J_0} \bar{t}^{\bar{J}_0}),
}
where the trace is over the ground states in the Ramond sector. For a
Landau-Ginzburg orbifold, this generalized Poincar\'{e} polynomial has been
calculated by Intriligator and Vafa~\refs{\rVS,\rIV}. The only difference with
respect to the Poincar\'{e} polynomial that we have discussed is that the
arguments of $P^{\rm tw}[R](g_a)$ and $P^{\rm inv}[R_i](g_b)$ in~\efourtw\ are
$t/\bar{t}$ and $t\bar{t}$ respectively instead of $t$. We may therefore
continue to work with the restricted Poincar\'{e} polynomial which only depends
on $t$. A criterion for mirror symmetry is then that the Poincar\'{e}
polynomials of the two models must be equal and that contributions from twisted
(untwisted) fields in one model should correspond to contributions from
untwisted (twisted) fields in the other.

A necessary condition for the elliptic genera of two $N = 2$ models to be equal
is that they coincide in the $\tau \rightarrow i \infty$ limit, \ie\ that the
two models have the same Poincar\'{e} polynomial. We will now show that in the
case of orbifolds of Landau-Ginzburg models with isomorphic groups of phase
symmetries, this condition is also
sufficient~\ft{In a more general case it may not be sufficient to merely
consider the $q\to 0$ limit~\rASY.}.
The following proof was first
given by Francesco and Yankielowicz~\rFY.
 Suppose that we have two Landau-Ginzburg
orbifolds with elliptic genera $Z[R/H]$ and $Z[\tilde{R} / \tilde{H}]$ such
that their Poincar\'{e} polynomials
agree (up to a sign which we will neglect in the discussion below).
It follows from~\emoduli\ that
the conformal anomalies $\hat{c}$ of the two models must be the same, and that
the function $f(z | \tau) = Z[R/H] / Z[\tilde{R} / \tilde{H}]$, where $z =
\gamma / L$,  transforms as follows:
\eqn\efourone{
\eqalign{
f(z + 1 | \tau) & =  f(z | \tau)  \cr
f(z + \tau | \tau) & =  f(z | \tau) \cr
f(z | \tau + 1) & =  f(z | \tau)  \cr
f(z / \tau | - 1 / \tau) & =  f(z | \tau) .
}}
The equality of the Poincar\'{e} polynomials means that $\lim_{\tau \rightarrow
i \infty} f(z | \tau) = 1$ for all $z$. Furthermore, from the properties of the
Jacobi $\Theta_1$ function, we have that $f(z | \tau)$ must be a meromorphic
function of $z$ for all $\tau$. The first two properties in~\efourone\ then
imply that $f$ may be written as~\rWW\
\eqn\eXXX{
f(z | \tau) = A \prod_{i = 1}^n \frac{\Theta_1(z - a_i(\tau) |
\tau)}{\Theta_1(z - b_i(\tau) | \tau)},
}
where $A$ is a constant, and $a_i(\tau)$ and $b_i(\tau)$ are the positions of
the zeros and poles in $z$ respectively. (The double periodicity in $z$ implies
in particular that the number of zeros and poles are equal.) The last two
properties in~\efourone\ give that
\eqn\efourtwo{
\eqalign{
\tau a_i(-1 / \tau) & =  a_{\sigma(i)}(\tau)
\quad  {\rm mod} \; 1 \quad {\rm mod} \; \tau \cr
\tau a_i(\tau + 1) & =  a_{\rho(i)}(\tau)
\quad  {\rm mod} \; 1 \quad {\rm mod} \; \tau ,
}}
where $\sigma$ and $\rho$ are some permutations of $\{ 1, \ldots , n \}$. For a
Landau-Ginzburg orbifold, the $a_i(\tau)$ and $b_i(\tau)$ are analytic
functions of $\tau$, but this need not be true for a more general $N = 2$
model~\rASY.
If $a_i(\tau)$ and $b_i(\tau)$ are analytic, one may show~\rFY, by using the
second equation in~\efourtwo\ $n! \, m$ times, combining with the first
equation in~\efourtwo\ and letting $m \rightarrow \infty$, that
\eqn\eXXX{
a_i(\tau) = \frac{1}{n!}(\alpha_i + \beta_i \tau)
}
for some constants $\alpha_i$ and $\beta_i$, which are chosen so that $0
\leq \alpha_i, \beta_i < n!$. Inserting this expression back into
\efourtwo\ we find that $\alpha_i, \beta_i \in \{ 0, 1, \ldots , n! -1
\}$. By a completely analogous argument, one can show~\rFY\ that
\eqn\eXXX{
b_i(\tau) = \frac{1}{n!}(\gamma_i + \delta_i \tau)
}
for some constants $\gamma_i, \delta_i \in \{ 0, 1, \ldots , n! -1 \}$. We may
now calculate the $\tau \rightarrow i \infty$ limit of $f(z | \tau)$ and
find~\rFY\
\eqn\eXXX{
A \frac{\prod_{\beta_i = 0} 2 \sin \pi (z - \alpha_i / n!) \prod_{\beta_i > 0}
(-i) q^{-\beta_i / (2 n!)} \exp( i \pi (z - \alpha_i))}{\prod_{\delta_i = 0} 2
\sin \pi (z - \gamma_i / n!) \prod_{\delta_i > 0} (-i) q^{-\delta_i / (2 n!)}
\exp (i \pi (z - \gamma_i))}.
}
A necessary condition for this expression to equal $1$ as $\tau \rightarrow i
\infty$ is that the real zeros (\ie\ $\alpha_i / n!$ for $\beta_i = 0$) cancel
against the real poles (\ie\ $\gamma_i / n!$ for $\delta_i = 0$).
Furthermore, we have already used the fact that the modular transformations
$(z, \tau) \rightarrow (z, \tau + 1)$ and $(z, \tau) \rightarrow (z / \tau, - 1
/ \tau)$ permute the zeros and the poles of $f(z | \tau)$ among themselves.
This means that the set of $(\alpha_i, \beta_i)$ is mapped into itself under
the transformations
\eqn\eXXX{
\eqalign{
(\alpha, \beta) & \rightarrow  (\alpha + \beta, \beta) \cr
(\alpha, \beta) & \rightarrow  (-\beta, \alpha) .
}}
For any $(\alpha, \beta)$ there exists a sequence of such transformations which
maps it into $({\rm GCD} (\alpha, \beta), 0)$, \ie\ a real zero. But we have
just argued that all real zeros must cancel against real poles. We must
therefore conclude that {\it all\/} zeros cancel against poles so that $f(z |
\tau) = A = 1$. Thus $Z[R/H] = Z[\tilde{R} / \tilde{H}]$, which concludes the
proof.

\newsec{Mirror symmetry for Landau-Ginzburg orbifolds}\noindent
In this section, we will discuss a very natural scenario for two
Landau-Ginzburg orbifolds to be mirror partners in the sense that their
elliptic genera are equal (up to a minus sign in the case when the number of
fields of the models is odd).

Suppose that we have two Landau-Ginzburg models, each with  $|N|$
superfields, the phase symmetry groups of
which are both isomorphic to the same abelian group $G$. We denote the sets of
representations as $R$ and $\tilde{R}$ respectively for the two models. In
general, we distinguish all quantities pertaining to the second model with a
tilde. We are interested in the situation in which the $H$ orbifold of the
first model is the mirror partner of the $\tilde{H}$ orbifold of the second
model for some subgroups $H$ and $\tilde{H}$ of $G$. This means that their
elliptic genera   should be equal up to a sign:
\eqn\enfivezero{
Z[R/H] = \pm Z[\tilde{R} / \tilde{H}].
}
We will propose a
natural way for two models to be each others mirror partners in this sense,
but to do so, we first need to discuss some aspects of abelian groups.

All irreducible representations of the abelian group $G$ are one-dimensional.
Given two irreducible representations we may construct a new irreducible
representation by taking their tensor product. Clearly, the set of irreducible
representations of $G$ form a group $G^*$ under the tensor product $\otimes$.
This group is in fact isomorphic to $G$ itself. Given a subgroup $H$ of $G$ we
define its dual as the subgroup $\tilde{H}$ of $G^*$ of representations on
which $H$ is trivially represented, \ie\ $\tilde{H}$ is the set of $R \in G^*$
such that $R(g) = 1$ for $g \in H$. In particular, the dual of $G$ itself is
the trivial subgroup of $G^*$ which only consists of the identity element, and
the dual of the trivial subgroup of $G$ is $G^*$. Clearly, if $H_1 \subset H_2$
then $\tilde{H_2} \subset \tilde{H_1}$. Similarly, we note that a
representation of the group $G^*$ has a natural interpretation as an element of
$G$. Therefore, given a subgroup $\tilde{H}$ of $G^*$ we may define its dual
$\tilde{\tilde{H}}$ as
 the subgroup of elements of $G$ which are trivially represented by all $R \in
\tilde{H}$. We see that $\tilde{\tilde{H}} = H$. The situation is thus
completely symmetric under interchange of $G$ and $G^*$.
As an alternative definition of the duality between $H$ and $\tilde H$
we have the following relation
\eqn\enfiveone{
\frac{1}{|H|} \sum_{g \in H} \tilde{g}(g) = \left\{ \eqalign{ 1 & \quad{\rm
for \;\;} \tilde{g} \in \tilde{H} \cr 0 & \quad{\rm otherwise} } \right.
}
and its partner obtained by changing the roles of $G$ and $G^*$. The
summand $\tilde g(g)$ is the function, defined on $G$, which
specifies the $G$ representation $\tilde g\in G^*$.

We now interpret our candidate mirror pair of Landau-Ginzburg models so that
the fields of the first model transform in the representations $R_i$ for $i \in
N$ under the symmetry group $G$, whereas the fields of the second model
transform in the representations $\tilde{R}_i$ for
$i \in \tilde{N}$ under the group $G^*$. As our notation suggests,
we will take the $H$ orbifold of the first model and the $\tilde{H}$ orbifold
of the second model, where $H$ and $\tilde{H}$ are dual subgroups.

For $g \in
G$ and $\tilde{g} \in G^*$ we now
define the (partial) Fourier transform of the
elliptic genus contributions as
\eqn\eXXX{
\hat{Z}[R](g, \tilde{g}) = \frac{1}{|G|} \sum_{g^\prime \in G} \tilde{g}
(g^\prime) Z[R](g, g^\prime),
}
which may be inverted by means of~\enfiveone\ to yield
\eqn\eXXX{
Z[R](g, g^\prime) = \sum_{\tilde{g} \in G^*} \tilde{g}^{-1}(g^\prime)
\hat{Z}[R](g,
\tilde{g}).
}
Inserting this in~\epartition\ and using~\enfiveone\ we get
\eqn\eXXX{
Z[R/H] = \sum_{h \in H} \sum_{\tilde{h} \in \tilde{H}} \hat{Z}[R](h,
\tilde{h}).
}
Analogously, we calculate
\eqn\eXXX{
Z[\tilde{R}/\tilde{H}] = \sum_{\tilde{h} \in \tilde{H}} \sum_{h \in H}
\hat{Z}[\tilde{R}](\tilde{h}, h).
}
A very natural way to fulfil~\enfivezero\ is then to require that
\eqn\enfivetwo{
\hat{Z}[R](g, \tilde{g}) = \pm \hat{Z}[\tilde{R}](\tilde{g}, g)
}
for $g \in G$ and $\tilde{g} \in G^*$.
We will say that  sets of representations $R$ and $\tilde R$ which
fulfil~\enfivetwo\ are conjugates of each other.
Note that this condition implies
\enfivezero\ for {\it any} $H$ and its dual $\tilde{H}$. Furthermore, given two
Landau-Ginzburg models $R_1$ and $R_2$ with symmetry groups isomorphic to $G_1$
and $G_2$ respectively, we may construct the product model $R = R_1 \times R_2$
with symmetry group $G \simeq G_1 \times G_2$. If now $R_1$ and $R_2$ are
conjugates to $\tilde{R}_1$ and $\tilde{R}_2$ respectively so that each
 pair satisfies~\enfivetwo, then the pair of product models $R = R_1
\times R_2$ and $\tilde{R} = \tilde{R}_1 \times \tilde{R}_2$ also satisfies
\enfivetwo. This means that any $H$ orbifold of $R$, for $H$ a subgroup of
$G \simeq G_1 \times G_2$, will be the mirror partner of the corresponding
$\tilde{H}$ orbifold of $\tilde{R}$, even if $H$ is not of the form $H_1 \times
H_2$ for any subgroups $H_1$ of $G_1$ and $H_2$ of $G_2$.

As we have seen in the previous section, in the case of Landau-Ginzburg
orbifolds it is sufficient to compare the $\tau \rightarrow i \infty$ limits of
elliptic genera, \ie\ the Poincar\'{e} polynomials, to establish their
equality for all $\tau$.
With the $\tau \rightarrow i \infty$ limit of $Z[R](g,
g^\prime)$ given by~\efourtw, the corresponding limit of $\hat{Z}[R](g,
\tilde{g})$ is
\eqn\ephat{
\hat{P}[R](g, \tilde{g}) = P^{\rm tw}[R](g) \frac{1}{|G|} \sum_{g^\prime \in G}
\tilde{g}(g^\prime) \prod_{R_i(g) = 1} P^{\rm inv}[R_i](g^\prime).
}
Similarly, we have
\eqn\eptihat{
\hat{P}[\tilde{R}](\tilde{g}, g) = P^{\rm tw}[\tilde{R}](\tilde{g})
\frac{1}{|G|} \sum_{\tilde{g}^\prime \in G^*} \tilde{g}^\prime(g)
\prod_{\tilde{R}_i(\tilde{g}) = 1} P^{\rm
inv}[\tilde{R}_i](\tilde{g}^\prime).
}
Our condition for mirror symmetry now reads
\eqn\enfivethree{
\hat{P}[R](g, \tilde{g}) = \pm \hat{P}[\tilde{R}](\tilde{g}, g).
}
If this condition is fulfilled, then obviously $P[R/H] = \pm
P[\tilde{R}/\tilde{H}]$ for any subgroup $H$. Our results from the last section
then imply that also~\enfivezero\ is satisfied. The natural way for this to
come about is that also~\enfivetwo\ holds. Although we have no proof, we
strongly believe that this is indeed always the case.

The obvious question is now how we may find a pair of Landau-Ginzburg
orbifolds such that~\enfivethree\ is obeyed. To answer this,
we must first introduce
some more notation. Let $s$ be a subset of the set $N$ that indexes the
fields. We will only be
interested in $s$ such that there is at least one element of $G$ which leaves
untwisted the $X_i$ for $i \in s$ and twists the remaining fields. We denote
the set of such $s$ as $S$, and henceforth we will always assume that $s \in
S$. For $g \in G$ we define $\sigma(g) \in S$ by the condition that $i \in
\sigma(g)$ if and only if $X_i$ is left untwisted by $g$. Next, we introduce
the subgroups $G_s$ of elements of $G$ that leave untwisted the fields $X_i$
for $i \in
s$. The remaining fields may be twisted or untwisted depending on which element
of $G_s$ we choose. The corresponding objects in the conjugate
 model are denoted
as $\tilde{s}$, $\tilde{S}$, $\tilde{\sigma}(\tilde{g})$ and $G^*_{\tilde{s}}$
respectively.

To find a pair of conjugate models,
we assume that there is a one-to-one map $\rho$ from
$S$ to $\tilde{S}$ such that $(-1)^{|s|} = (-1)^{|N|-|\rho(s)|}$ and $G_s
\simeq \widetilde{G^*_{\rho(s)}}$ for $s \in S$. Here, $|s|$ denotes the number
of elements in $s$, and as usual the tilde over $G^*_{\rho(s)}$
denotes the dual group. Furthermore, we demand that
\eqn\efivefive{
\prod_{i \in s} P^{\rm inv}[R_i](g) = \sum_{G^*_{\tilde{s}} \subseteq
\widetilde{G_s}} (-1)^{|N|-|\tilde{s}|}
\sum_{\tilde{g} \in G^*_{\tilde{s}}}
\tilde{g}(g^{-1}) P^{\rm tw}[\tilde{R}](\tilde{g}).
}
Our notation means that the first sum runs over all $\tilde{s} \in \tilde{S}$
such that
$G^*_{\tilde{s}} \subseteq \widetilde{G_s}$. We also postulate the
corresponding relation with the roles of the two models interchanged. At this
point, the conditions that we have imposed may seem rather {\it ad hoc}. Our
main justification is that they cover all cases of mirror symmetry between
Landau-Ginzburg orbifolds that we know of. We have performed some limited
computer searches, which support the hypothesis that this is indeed the general
mechanism for mirror symmetry between Landau-Ginzburg orbifolds.

To verify that these conditions are sufficient for $R$ and $\tilde R$
to be conjugate, we
insert~\efivefive\ in~\ephat\ and~\eptihat\ and use~\enfiveone\ to get
\eqn\eXXX{
\eqalign{
\hat{P}[R](g, \tilde{g})  = & P^{\rm tw}[R](g) \frac{1}{|G|} \sum_{g^\prime
\in G} \tilde{g}(g^\prime) \prod_{i \in \sigma(g)} P^{\rm
inv}[R_i](g^\prime) \cr
 = & P^{\rm tw}[R](g) \, P^{\rm tw}[\tilde{R}](\tilde{g}) \sum_{\tilde{g} \in
G^*_{\tilde{s}} \subseteq \widetilde{G_{\sigma(g)}}} (-1)^{|N|-|\tilde{s}|} \cr
\hat{P}[\tilde{R}](\tilde{g}, g) = & P^{\rm tw}[R](g) \, P^{\rm
tw}[\tilde{R}](\tilde{g}) \sum_{g \in G_s \subseteq
\widetilde{G^*_{\tilde{\sigma}(\tilde{g})}}} (-1)^{|N|-|s|}.
}}
The notation means that the sum in the last equation runs over all $s
\in S$ such
that $g \in G_s \subseteq \widetilde{G^*_{\tilde{\sigma}(\tilde{g})}}$. But $g$
leaves $X_i$ untwisted if and only if $i \in \sigma(g)$. Therefore, $g \in G_s$
implies that $s$ must be a subset of $\sigma(g)$, and consequently
$G_{\sigma(g)}
\subseteq G_s$. Conversely, $G_{\sigma(g)} \subseteq G_s$ of course implies
that $g \in G_s$. The sum thus runs over all $s \in S$ such that $G_{\sigma(g)}
\subseteq G_s \subseteq \widetilde{G^*_{\tilde{\sigma}(\tilde{g})}}$. By an
analogous argument, we find that the sum in the first equation runs over all
$\tilde{s} \in \tilde{S}$ such that $G^*_{\tilde{\sigma}(\tilde{g})} \subseteq
G^*_{\tilde{s}} \subseteq \widetilde{G_{\sigma(g)}}$. By taking the dual and
changing variables according to $\tilde{s} = \rho(s)$ we see that the sums in
the two equations run over the same set. Furthermore, $(-1)^{|N|-|\tilde{s}|} =
(-1)^{|N|} (-1)^{|N|-|s|}$. It follows that
\eqn\eXXX{
\hat{P}[R](g, \tilde{g}) = (-1)^{|N|} \hat{P}[\tilde{R}](\tilde{g}, g),
}
which proves~\enfivethree. We see that in this way of
implementing mirror symmetry, contributions from twisted fields in one model
correspond to contributions from untwisted fields in the other model and vice
versa, just as it should be for a mirror pair.

Finally, we make the following interesting observation which will be useful
in the next section. By using~\epinv\
the left-hand side of~\efivefive\ picks up
no phase
(or a minus sign depending on whether the number of elements in $s$ is odd or
even) if we let $t$ encircle the origin counterclockwise and simultaneously
change $g \rightarrow g q^{-1}$. (Here, $q$ is the element of $G$ for which
$R_i(q) = \exp i 2 \pi q_i$ for $i \in N$.) For the right-hand side of
\efivefive\ to share this property, it is natural in view of~\eptw\ to require
that
\eqn\efiveeight{
\prod_{i \in \tilde{N}} \tilde{R}_i = q.
}
This condition makes sense since the $\tilde{R}_i$, being
representations of the group $G^*$ of representations of $G$, may be
interpreted as elements of $G$. Of course, we should also impose the
corresponding condition with the roles of the two models interchanged.

\newsec{Two examples}
\noindent
In this section, we will describe two classes of solutions, first proposed
in~\rBH, to the conditions
\efivefive. Both of these can be seen as generalizations of
Landau-Ginzburg analogs of the $(2,2)$
minimal models for which mirror symmetry was first discovered~\rGP.
Let us here once again stress that the $N=2$ minimal models
(including products and/or quotients thereof) are the only
theories for which mirror symmetry has been rigorously proven. In terms of
the Landau-Ginzburg models of Fermat type it has been {\it conjectured}
that orbifolds of a Fermat potential come in mirror pairs. This conjecture is
based on studies of the spectra of Landau-Ginzburg vacua~\refs{\rMPR-\rAR}
and their orbifolds~\rMax\ as well as more detailed investigations
of the moduli space of particular $\hat c=3$ theories~\refs{\rCdGP,\rAGM}.
In particular the work in~\refs{\rBH,\rMax} as well as recent advances
in terms of toric geometry~\refs{\rBatdual,\rCOK} indicate that mirror
symmetry must hold for a much larger class of theories than the minimal
models. As was noted in the previous
section, we may construct new solutions by taking the product of
old ones. We conjecture that by taking products of the models we will describe
in this section, one may in fact construct all solutions to the
conditions~\efivefive.

\subsec{The linear chain}
\noindent
The first class of models is given by a potential of the form~\rArnold\
\eqn\esixone{
W = X_1^{\al_1} + X_1 X_2^{\al_2} + \ldots + X_{N-1} X_N^{\al_N}.
}
Models of this type was extensively studied as part of the classification of
Landau-Ginzburg vacua with $\hat c=3$~\refs{\rMPR-\rAR}.
The group of phase symmetries is isomorphic to $G \simeq \ZZ_D$, where $D =
\al_1
\ldots \al_N$. The field $X_N$ transforms under this group in a representation
$R_N \in G^*$, which we take to be of order $D$, so that the whole
phase symmetry group $G$ is non-trivially represented rather than some
subgroup. This $R_N$ then generates the whole group $G^*$ of representations of
$G$, and $R_N(g) = 1$ if and only if $g$ is the identity element of $G$. In
other words, if we parametrized $G \simeq \ZZ_D$ by an integer $g$ defined
modulo
$D$, then $R_N(g) = \exp (i 2 \pi r_N g / D)$ for some integer $r_N$ defined
modulo $D$. The requirement that the representation $R_N$ be of order $D$ is
tantamount to $r_N$ and $D$ being relatively prime. The representations of the
other fields are determined by the requirement that each term in the potential
$W$ be invariant under $G$. We thus get
\eqn\esixfive{
R_{i-1} = R_i^{\otimes (- \al_i)} \;\; {\rm for} \;\; 2 \leq i \leq N.
}
Notice that this means that $R_1^{\otimes \al_1}$ is the trivial
representation, so the first term in $W$ is indeed $G$ invariant. Furthermore,
the $U(1)$ charge of each term in $W$ should equal 1. This means that the
$U(1)$ charges $q_i$ of the $X_i$ are given by the equation
\eqn\esixsix{
\sum_{j=1}^N \alpha_{ij} q_j = 1 {\;\; \rm for \;\;} 1 \leq i \leq N,
}
where the matrix $\alpha$ is given by
\eqn\eXXX{
\alpha_{ij} = \left\{ \eqalign{
\alpha_i  \quad & {\rm for} \quad i = j \cr
1 \quad & {\rm for} \quad i = j + 1 \cr
0 \quad & {\rm otherwise} }
\right. .
}
We see from~\esixfive\ that if the field $X_i$ is left untwisted by a given
element $g \in G$, then the same is true for $X_j$ for $1 \leq j \leq i$. The
set $S$ of subsets of fields that we introduced in the last section may
therefore be identified with the set $S = \{0, \ldots, N\}$. We interpret this
so that for any $s \in S$ we have $i \in s$ exactly for $1 \leq i \leq s$.
Clearly, the number of elements of $s$ is $|s| = s$.
Although this notation is
slightly abusive, it should be clear from the context whether $s$ stands for a
number or a set of fields. The
subgroup of elements of $G$ that leave $X_i$ untwisted for $1 \leq i \leq s$ is
\eqn\eXXX{
G_s = \{g: \, g^{\al_{s+1} \ldots \al_N} = e \},
}
where $e$ is the identity element of $G$.

The conjugate partner of the potential $W$ is a potential of the same type but
with the order of the exponents reversed~\rBH, \ie\
\eqn\esixtwo{
\tilde{W} = \tilde{X}_1^{\tilde{\al}_1} + \tilde{X}_1
\tilde{X}_2^{\tilde{\al}_2} + \ldots + \tilde{X}_{N-1}
\tilde{X}_N^{\tilde{\al}_N},
}
where the exponents are given by
\eqn\eXXX{
\tilde{\al}_i = \al_{N+1-i} \;\;  {\rm for} \;\; 1 \leq i \leq N.
}
The group of phase symmetries is obviously isomorphic to $G^* \simeq \ZZ_D$,
and
our previous considerations about the representations and $U(1)$ charges of the
fields of $W$ apply to $\tilde{W}$ as well. In particular we have that
$\tilde{R}_N \in G$ is of order $D$ and that the representations of the other
fields are determined recursively by
\eqn\esixten{
\tilde{R}_{i-1} = \tilde{R}_i^{(- \tilde{\al}_i)}  = \tilde{R}_i^{(-
\al_{N+1-i})} \;\; {\rm for} \;\; 2 \leq i \leq N.
}
It may seem that there is some freedom in the choice of $R_N \in G^*$ and
$\tilde{R}_N \in G$, but acting with $R_N$ on equation~\efiveeight\ and
expressing the $\tilde{R}_i$ for $1 \leq i \leq N-1$ in terms of $\tilde{R}_N$
by means of~\esixten\ we get
\eqn\eXXX{
 R_N(\tilde{R}_N^r) = R_N(q),
}
where
\eqn\eXXX{
r = 1 - \tilde{\al}_N + \ldots + (-1)^{N-1} \tilde{\al}_N \ldots \tilde{\al}_2
= (-1)^{N-1} D q_N
}
so that
\eqn\eXXX{
R_N(\tilde{R}_N) = \exp (\frac{i 2 \pi (-1)^{N-1}}{D}).
}
For each $R_N \in G^*$ of order $D$ there is exactly one solution to this
condition, and the $\tilde{R}_N \in G$ thus determined is also of order $D$. It
then follows from~\esixfive\ and~\esixten\ that
\eqn\esixtw{
R_i (\tilde{R}_{N-j}) = \left\{ \eqalign{ 1\qquad & \quad
{\rm for} \quad i \leq j\cr
\exp( \frac{i 2 \pi (-1)^{j+1-i}}{\al_{j+1} \ldots \al_i})
& \quad {\rm for} \quad i >j
} \right. .
}
Finally, the map $\rho$ from $S$ to $\tilde{S}$ is given by
\eqn\eXXX{
\rho (s) = N - s \;\; {\rm for} \;\; 0 \leq s \leq N.
}

To prove that the sets of representations $R$ and $\tilde R$
corresponding to $W$ and $\tilde W$ respectively are conjugate
we
must verify that condition~\efivefive\ is fulfilled. In our case it amounts to
\eqn\esixthree{
\prod_{i = 1}^s P^{\rm inv}[R_i](g) = \sum_{\tilde{s} = N - s}^N (-1)^{N -
\tilde{s}} \sum_{\tilde{g} \in G^*_{\tilde{s}}} P^{\rm
tw}[\tilde{R}](\tilde{g}) \, \tilde{g}(g^{-1}).
}
We will prove~\esixthree\ by induction over $s$, but we will have to
distinguish between the cases when $s$ is odd and $s$ is even. We should thus
verify the statement for the cases when $s = 0$ and $s = 1$ separately, and
show that
\eqn\esixfour{
\eqalign{
&\prod_{i = 1}^s P^{\rm inv}[R_i](g) - \prod_{i = 1}^{s-2} P^{\rm inv}[R_i](g)
=
\cr
&(-1)^s \sum_{\tilde{g} \in G^*_{N-s}} P^{\rm tw}
[\tilde{R}](\tilde{g}) \, \tilde{g}(g^{-1}) + (-1)^{s-1} \sum_{\tilde{g} \in
G^*_{N-s+1}} P^{\rm tw} [\tilde{R}](\tilde{g}) \, \tilde{g}(g^{-1})
}}
for $2 \leq s \leq N$.
The left-hand side of~\esixfour\ equals
\eqn\eXXX{\hbox{}\mkern30mu
\eqalign{
 lhs=
\prod_{i = 1}^s & t^{-1/2} (1 - t^{q_i} R_i(g))^{-1} \prod_{i = 1}^{s-2}
(t^{q_i} R_i(g) - t) \cr
\times &\Big( (t^{q_{s{-}1}} R_{s{-}1}(g) {-} t) (t^{q_s} R_s(g) {-} t) - t
(1 {-} t^{q_{s{-}1}} R_{s{-}1}(g)) (1 {-} t^{q_s} R_s(g)) \Big) \cr
= &\prod_{i = 1}^s t^{-1/2} (1 - t^{q_i} R_i(g))^{-1}\cr
 \times &\prod_{i = 1}^{s{-}2} (t^{q_i} R_i(g) {-} t)
  t^{q_s{-}1} R_{s{-}1}(g) (t^{q_s} R_s(g) {-} t^{1 {-} q_{s{-}1}}
R_{s{-}1}^{{-}1}) (1 {-} t).
}}
Using~\esixfive\ and~\esixsix\ for $1 \leq i \leq N$ and $R_1^{\al_1}(g) =
1$, we may write this as
\eqn\esixseven{
\eqalign{
lhs=
&\prod_{i = 1}^s  t^{{-}1/2} (1 {-} t^{q_i} R_i(g))^{-1} \prod_{i = 1}^{s{-}1}
(t^{q_i} R_i(g) {-} t^{(1{+}\al_i) q_i} R_i^{1{+}\al_i}(g)) \,
(t^{q_s} R_s(g) {-}
t^{\al_s q_s} R_s^{\al_s}(g)) \cr
= & t^{-s/2} \sum_{p_1 = 1}^{\al_1} \ldots \sum_{p_{s{-}1} = 1}^{\al_{s{-}1}}
 \sum_{p_s = 1}^{-1 {+} \al_s} t^{p_1 q_1 {+} \ldots {+} p_s q_s}
(R_1^{\otimes p_1}
\otimes \ldots \otimes R_s^{\otimes p_s}) (g) .
}}
Inserting
\eqn\eXXX{
P^{\rm tw}[\tilde{R}](\tilde{g}^{-1}) = (-t^{-1/2})^{{\rm
tw}[\tilde{R}](\tilde{g})}
\, t^{\sum_{i = 1}^N [\tilde{\theta}_i(\tilde{g})]}
}
in the right-hand side of~\esixfour\ and using the fact that $G^*_{N-s} -
G^*_{N-s+1}$ equals the set $\tilde{T}_s$ of elements that twist the fields
$\tilde{X}_i$ for $N-s+1 \leq i \leq N$ and leave
the remaining fields untwisted, we get for the right-hand side of~\esixfour\
\eqn\esixeight{
t^{-s/2} \sum_{\tilde{g} \in \tilde{T}_{N-s}}
t^{[\tilde{\theta}_{N-s+1}(\tilde{g})] + \ldots +
[\tilde{\theta}_N(\tilde{g})]} \, \tilde{g}(g).
}
We should therefore prove that~\esixseven\ equals~\esixeight\ for $2 \leq s
\leq N$. Before we do that, however, we note that we should also
verify~\esixthree\ for the cases $s = 0$ and $s = 1$. For $s = 0$ the left-hand
side of~\esixthree\ is trivially $1$, which equals the right-hand side since
the
sum over $\tilde{g}$ runs over $G^*_N$, which only consists of the trivial
representation of $G$. For the case $s = 1$ we use once more that
$R_1^{\al_1}(g) = 1$ and $\al_1 q_1 = 1$ to see that the left-hand side of
\esixthree\ equals~\esixseven. Furthermore, for $s = 1$, the right-hand side
of~\esixthree\ equals~\esixeight. We are therefore ready if we can prove
that~\esixseven\ and~\esixeight\ are equal for $1 \leq s \leq N$, \ie\ that
\eqn\esixnine{
\eqalign{
\sum_{p_1 = 1}^{\al_1} \ldots \sum_{p_{s-1} = 1}^{\al_{s-1}}  &\sum_{p_s =
1}^{-1 + \al_s} t^{p_1 q_1 + \ldots + p_s q_s} (R_1^{\otimes p_1} \otimes
\ldots \otimes R_s^{\otimes p_s}) (g) \cr
= & \sum_{\tilde{g} \in \tilde{T}_{N-s}}
t^{[\tilde{\theta}_{N-s+1}(\tilde{g})] + \ldots +
[\tilde{\theta}_N(\tilde{g})]} \, \tilde{g}(g).}
}

The proof of~\esixnine\ consists of two parts. First, we should show that
$R_1^{\otimes p_1} \otimes \ldots \otimes R_s^{\otimes p_s}$ goes through all
representations of $G$ in $\tilde{T}_{N-s}$ as $p_i$ take the values $1 \leq
p_i \leq \al_i$ for $1 \leq i \leq s-1$ and $1 \leq p_s \leq -1 + \al_s$.
Furthermore, we should show that
\eqn\esixth{
p_1 q_1 + \ldots + p_s q_s = [\tilde{\theta}_{N-s+1}(\tilde{g})] + \ldots +
[\tilde{\theta}_N(\tilde{g})]
}
when $\tilde{g} = R_1^{\otimes p_1} \otimes \ldots \otimes R_s^{\otimes p_s}
\in \tilde{T}_{N-s}$.
The first part follows by using~\esixtw. To prove~\esixth, we use~\esixsix\ to
rewrite the left-hand side as
\eqn\eXXX{
p_1 q_1 + \ldots + p_s q_s = k_1 + \ldots + k_s,
}
where the $k_i$ are given by the equation
\eqn\esixtwefo{
\sum_{i=1}^s k_i \alpha_{ij} = p_j {\;\; \rm for \;\;} 1 \leq j \leq s.
}
But from~\esixtw\ it follows that
\eqn\eXXX{
\al_j [\tilde{\theta}_{N-j+1} (\tilde{g})] = p_j - [\tilde{\theta}_{N-(j+1)+1}
(\tilde{g})]
}
or, equivalently,
\eqn\esixtwefi{
\sum_{i=1}^s [\tilde{\theta}_{N-i+1} (\tilde{g})] \alpha_{ij} = p_j.
}
Comparing~\esixtwefo\ and~\esixtwefi\ we see that $\tilde{\theta}_{N-i+1}
(\tilde{g}) = k_i$ for $1 \leq i \leq s$, which proves~\esixth\ and
therefore \esixnine. Since we have now proved~\esixthree\ for $s=0$ and
$s=1$ and~\esixfour\ for $2 \leq s \leq N$, \esixthree\ for $0 \leq s \leq
N$ follows by induction over $s$. This is equivalent to~\efivefive. The
argument may be repeted after exchanging the roles of the potentials $W$ and
$\tilde{W}$. This concludes the proof that
 $R$ and $\tilde R$ indeed are conjugate representations.

The subgroups of $G \simeq \ZZ_D$ are isomorphic to $H \simeq \ZZ_m$ for some
$m$
which divides $D$. The dual subgroup of $G^*$ is then isomorphic to $\tilde{H}
\simeq \ZZ_{\tilde{m}}$, where $\tilde{m} = D/m$. The general arguments of the
previous section now show that the $H$ orbifold and the $\tilde{H}$ orbifold of
the Landau-Ginzburg models with potentials $W$ and $\tilde{W}$ are mirror
partners.

\subsec{The Loop}
\noindent
Our second example is in many respects similar to the first one, and our
treatment of it will largely parallel the above discussion. This class of
models is given by a potential of the form
\eqn\eXXX{
W = X_N X_1^{\al_1} + X_1 X_2^{\al_2} + \ldots + X_{N-1} X_N^{\al_N}.
}
The cyclical nature of this potential makes it natural to take the variable
$i$, which indexes the fields, to be defined modulo $N$, \ie\ $i \in \ZZ_N$.
The
group of phase symmetries is isomorphic to $G \simeq \ZZ_D$, where $D = \al_1
\ldots \al_N + (-1)^{N-1}$. As usual, we demand that each term in $W$ be
invariant under $G$ and of $U(1)$ charge $1$. For the representations $R_i$ and
the $U(1)$ charges $q_i$ of the fields $X_i$ this means that
\eqn\esixfo{
R_{i-1} = R_i^{\otimes (- \al_i)}
}
and
\eqn\esixse{
\sum_{j \in \ZZ_N} \alpha_{ij} q_j = 1
}
for $i \in \ZZ_N$. Here, the matrix $\alpha$ is given by
\eqn\eXXX{
\alpha_{ij} = \left\{ \eqalign{
\alpha_i \quad & {\rm for} \quad i = j \cr
1 \quad& {\rm for} \quad i = j + 1 \cr
0 \quad& {\rm otherwise} }
\right. .
}
We see from~\esixfo\ that all the representations $R_i$ are of the same
order, which we take to equal $D$. Thus $R_i(g) = 1$ if and only if $g$ is the
trivial element of $G$. The set of subsets of fields $S$ of the previous
section may therefore be identified with $S = \{ 0, N \}$, so that $s = N$
includes all fields whereas $s = 0$ is empty. The subgroup of elements of $G$
that leave all fields untwisted is the trivial subgroup $G_N$, which only
consists of the identity element of $G$. The subgroup $G_0$ is of course $G$
itself.

The conjugate
 partner of this potential is a potential of the same type but with
the order of the exponents $\al_i$ reversed~\rBH, \ie\
\eqn\eXXX{
\tilde{W} = \tilde{X}_N \tilde{X}_1^{\tilde{\al}_1} + \tilde{X}_1
\tilde{X}_2^{\tilde{\al}_2} + \ldots + \tilde{X}_{N-1}
\tilde{X}_{N}^{\tilde{\al}_N},
}
where the exponents are given by
\eqn\eXXX{
\tilde{\al}_i = \al_{N+1-i} \;\; {\rm for \;\;} i \in \ZZ_N.
}
The group of phase symmetries is isomorphic to $G^* \simeq \ZZ_D$, and
precisely
as for the potential $W$ we take the fields $\tilde{X}_i$ to
transform in representations of order $D$ that fulfil
\eqn\esixfi{
\tilde{R}_{i - 1} = \tilde{R}_i^{(- \tilde{\al}_i)}.
}
To get a relation between the $R_i$ and the $\tilde{R}_j$, we act on
equation~\efiveeight\ with $R_i$ and express the $\tilde{R}_j$ in terms of
$\tilde{R}_{N-i}$ for some $i \in \ZZ_N$ using~\esixfi. We get
\eqn\eXXX{
R_i (\tilde{R}_{N-i}^r) = R_i (q),
}
where
\eqn\eXXX{
r = 1 - \tilde{\al}_{N-i} + \ldots + (-1)^{N-1} \tilde{\al}_{N-i}
\tilde{\al}_{N-i-1} \ldots \tilde{\al}_{N+2-i} = (-1)^{N-1} D q_i
}
so that
\eqn\esixni{
R_i (\tilde{R}_{N-i}) = \exp (\frac{i 2 \pi (-1)^{N-1}}{D}).
}
For each set of $R_i \in G^*$ of order $D$ there is exactly one solution to
this condition, and the $\tilde{R}_i \in G$ are also of order $D$.
Finally, the
map $\rho$ from $S$ to $\tilde{S}$ is given by
\eqn\eXXX{
\eqalign{
\rho(N)  = & 0 \cr
\rho(0)  = & N.
}}

To prove that the potentials $W$ and $\tilde{W}$ constitute a conjugate
 pair, we
must again verify that~\efivefive\ is fulfilled. For the
present model, this condition is non-trivial only for $s = N$, where it reads
\eqn\eXXX{
\prod_{i \in Z_N} P^{\rm inv}[R_i](g) = (-1)^N \sum_{\tilde{g} \in G^*_0}
P^{\rm tw}[\tilde{R}](\tilde{g}) \, \tilde{g}(g^{-1}) + \sum_{\tilde{g} \in
G^*_N}
P^{\rm tw}[\tilde{R}](\tilde{g}) \, \tilde{g}(g^{-1}).
}
Using~\esixfo, \esixse\ and the cyclical property of $i$, the
left-hand side of this equation may be rewritten as
\eqn\esixtwe{
\eqalign{
\prod_{i \in Z_N} P^{\rm inv}[R_i](g) & =  \prod_{i = 1}^N t^{-1/2} (t^{q_i}
R_i(g) - t) (1 - t^{q_i} R_i(g))^{-1} \cr
& =  t^{-N/2} \prod_{i = 1}^N (t^{q_i} R_i(g) - (t^{q_i} R_i(g))^{1 + \al_i})
(1 - t^{q_i} R_i(g))^{-1} \cr
& =  t^{-N/2} \sum_{p_1 = 1}^{\al_1} \ldots \sum_{p_N = 1}^{\al_N} t^{p_1 q_1
+ \ldots + p_N q_N} (R_1^{\otimes p_1} \otimes \ldots \otimes R_N^{\otimes
p_N})(g),
}}
whereas the right-hand side equals
\eqn\esixtweo{
\eqalign{
(-1)^N & \sum_{\tilde{g} \in G^*} (-t^{-1/2})^{{\rm tw}[\tilde{R}](\tilde{g})}
\, t^{\sum_{i = 1}^N [\tilde{\theta}_i(\tilde{g})]}
\, \tilde{g}(g) + 1 \cr
= t^{-N/2} & \sum_{\tilde{g} \in \tilde{T}_0} t^{\sum_{i
= 1}^N [\tilde{\theta}_i(\tilde{g})]} + (-1)^N + 1.}
}
Here, $\tilde T_0=G^*-\{e\}$ is the set of elements of $G^*$ that twist all
fields.
Using~\esixfo\ we may write $R_1^{\otimes p_1} \otimes \ldots \otimes
R_N^{\otimes p_N} = R_j^{\otimes p}$ for some arbitrary $j \in \ZZ_N$. As the
$p_i$ take the values $1 \leq p_i \leq \al_i$ for $1 \leq i \leq N$, $p$ takes
the values $0 \leq -p \leq \al_1 \ldots \al_N - 1 = D$ for $N$ even and $1 \leq
p \leq \al_1 \ldots \al_N = D-1$ for $N$ odd. We see that for $N$ odd, $R_N^p$
runs over all elements of $\tilde{T}_0$, whereas for $N$ even we also get the
identity element of $G^*$ twice. To prove the equality of~\esixtwe\ and
\esixtweo, we need now only check that
\eqn\esixei{
p_1 q_1 + \ldots + p_N q_N = [\tilde{\theta}_1(\tilde{g})] + \ldots +
[\tilde{\theta}_N(\tilde{g})]
}
when $\tilde{g} = R_1^{\otimes p_1} \otimes \ldots \otimes R_N^{\otimes p_N}
\in \tilde{T}_0$. Using~\esixse, we may rewrite the left-hand side as
\eqn\eXXX{
p_1 q_1 + \ldots + p_N q_N = k_1 + \ldots + k_N,
}
where the $k_i$ are given by the equation
\eqn\eXXX{
\sum_{i \in \ZZ_N} k_i \alpha_{ij} = p_j {\;\; \rm for \;\;} j \in \ZZ_N.
}
But by using~\esixfo, \esixfi\ and~\esixni\ one may show that
the $[\tilde{\theta}_{N+1-i}(\tilde{g})]$ fulfil
\eqn\eXXX{
\sum_{i \in \ZZ_N} [\tilde{\theta}_{N+1-i}(\tilde{g})] \alpha_{ij} = p_j {\;\;
\rm for \;\;} j \in \ZZ_N,
}
so $k_i = [\tilde{\theta}_{N+1-i}(\tilde{g})]$ for $i \in \ZZ_N$, which proves
\esixei\ and therefore~\efivefive. As before, we may repeat the argument after
exchanging
the roles of the potentials $W$ and $\tilde{W}$, and therefore the potentials
$W$ and $\tilde{W}$ indeed constitute a conjugate pair.

As for the first example, we may construct mirror pairs of orbifolds of these
models by taking subgroups $H \simeq \ZZ_m$ and
$\tilde{H} \simeq \ZZ_{\tilde{m}}$
for $m \tilde{m} = D$.

\vskip 5mm

{\bf Acknowledgements}:
The authors acknowledge useful discussions with T.~H\"ubsch and E.~Witten.
P.B. was supported by DOE grant DE-FG02-90ER40542 and
M.H. was supported by the Swedish Natural Science Research Council (NFR).

\vfill \eject

\listrefs

\bye